\documentclass[12pt]{article}

\usepackage[T1]{fontenc}
\usepackage[utf8]{inputenc} 

\usepackage{mathtools}
\usepackage{amscd,amsfonts,amssymb,amsthm,amscd,amsmath}
\usepackage{bm}
\usepackage{caption}
\usepackage{cite}
\usepackage[mathscr]{eucal} 
 \usepackage{latexsym}
\usepackage{mathptmx}
\usepackage{mathrsfs} 
\usepackage{mathtools}

\usepackage{setspace}
\newlength{\dinwidth}
\newlength{\dinmargin}
\newtheorem{theorem}{Theorem}

\numberwithin{equation}{section}

\def\idty{{\leavevmode\hbox{\rm 1\kern -.3em I}}}

\def\As{{\cal A}}

\def\Cs{{\cal C}}

\def\Hs{{\cal H}}

\def\Ls{{\cal L}}

\def\Os{{\cal O}}
\def\Ps{{\cal P}}

\def\Pid{{\Ps_+ ^{\uparrow}}}

\def\idty{{\leavevmode\hbox{\rm 1\kern -.3em I}}}
\def\RR{{\mathbb R}}

\def\INO{{{\mathbb N}_0}}

\def\CC{{\mathbb C}}

\def\In{{\textnormal{\tiny in}}}
\def\Out{{\textnormal{\tiny out}}}

\def\Pid{{\Ps_+ ^{\uparrow}}}

\newcommand{\ie}{{\it i.e.\ }}
\newcommand{\eg}{{\it e.g.\ }}
\newcommand{\etc}{{\it etc}}

\begin{document}

\title{Scattering in relativistic quantum field theory: \\
basic concepts, tools, and results} 
\author{Detlev Buchholz$\, {}^a$ \ and \ Wojciech Dybalski$\, {}^b$}
\date{
  \normalsize $\, {}^a$ Mathematisches Institut, Universit\"at
  G\"ottingen \\[0.5mm]
  Bunsenstr~3-5, \ 37073 G\"ottingen, Germany \\[1.5mm]
  \normalsize $\, {}^b$ Faculty of Mathematics and Computer Science, 
  Adam Mickiewicz University \\[0.5mm]
  ul.~Uniwersytetu Pozna\'nskiego 4, 61-614 Pozna\'n, Poland}

\maketitle 

\abstract{\noindent
  We provide an overview of basic concepts,
  tools, and results of quantum field theoretical scattering theory.
  This article is prepared for the second edition of the
  Encyclopedia of Mathematical Physics, edited by M.~Bojowald
  and R.J.~Szabo, to be published by Elsevier.}

\section{Physical Motivation and Mathematical Setting} \label{motivation}

     The primary connection of relativistic quantum field theory to
experimental physics is through scattering theory, \ie the theory of
the collision of elementary (or compound) particles.  It is therefore
a central topic in quantum field theory and has attracted the
attention of leading mathematical physicists. Although a
great deal of progress has been made in the mathematically rigorous
understanding of the subject, there are important matters which are
still unclear, some of which will be {outlined below.
References to articles in this encyclopedia have an E in
front of their number. Other references are given by their numbers only.}
  
     In the paradigmatic scattering experiment, several particles,
which are initially sufficiently distant from each other, {so that 
the idealization that they do not interact is physically
reasonable, approach each other and collide in a region of
microscopic extent}. The products of this collision then fly apart
until they are sufficiently well separated that the approximation of
non-interaction is again reasonable. The initial and final states of
the objects in the scattering experiment are therefore to be modeled
by states of non-interacting, \ie free, fields, which are
mathematically represented on Fock
space. Typically, what is measured in such experiments is the
probability distribution (cross section) for the transitions from a
specified state of the incoming particles to a specified state of the
outgoing particles.

     It should be mentioned that until the late 1950's, the scattering
theory of relativistic quantum particles relied upon ideas from
non-relativistic quantum mechanical scattering theory (interaction
representation, adiabatic limit, \etc), which were invalid in the
relativistic context. Only with the advent of axiomatic quantum field
theory did it become possible to properly formulate the concepts and
mathematical techniques which will be outlined here.

     Scattering theory can be rigorously formulated either in the
     context of quantum fields satisfying the Wightman axioms
     \cite{StWi, 1Enc}
or in terms of local algebras satisfying the
Haag-Kastler-Araki axioms \cite{Haag},
{cf.\ also \cite{2Enc}}. In brief,
the relation between these two settings may be described
as follows: In the Wightman setting, the theory is formulated
in terms of operator valued distributions $\phi$ on
Minkowski space, the quantum fields, which act 
on the physical state space. These 
fields, integrated with test functions $f$ having support 
in a given region $\Os$ of space-time,\footnote{We restrict 
ourselves to four-dimensional Minkowski space.
For theories in lower dimensions, cf.~\cite{3Enc}.}  
$\phi(f) = \int \! d^{4} \! x \; f(x) \phi (x)$, form under the 
operations of addition, multiplication and hermitian 
conjugation a polynomial *-algebra $\Ps (\Os)$ of unbounded operators.
In the Haag-Kastler-Araki setting one proceeds 
from these algebras to algebras $\As  (\Os)$ 
of bounded operators which, roughly speaking, are formed by the   
bounded functions $A$ of the operators $\phi(f)$. This step 
requires some mathematical care, but these subtleties will not 
be discussed here. As the statements and proofs
of the results in these two frameworks differ only in technical details, 
the theory is presented here in the more convenient setting of algebras of 
bounded operators (C*-algebras).

     Central to the theory is the notion of a particle, which, in
fact, is a quite complex concept, the full nature of which is not
completely understood, cf.\ below. In order to maintain the focus on 
the essential points, we consider in the subsequent sections primarily 
a single massive particle of integer spin $s$, \ie a Boson. 
In standard scattering theory based upon Wigner's characterization, 
this particle is simply identified with an irreducible unitary 
representation $U_1$ of the identity component $\Pid$ of the Poincar\'e group
with spin $s$ and mass $m > 0$. The Hilbert space $\Hs_1$ upon which 
$U_1(\Pid)$ acts is called the one-particle space and determines 
the possible states of a single particle, alone in the universe. 
Assuming that configurations of several such  
particles do not interact, one can
proceed by a standard construction to a Fock space
describing freely propagating multiple particle states, 
$$\Hs_F = \bigoplus_{n \in \INO } \, \Hs_n \, ,$$
where $\Hs_0 = \CC$ {represents the vacuum} and $\Hs_n$
is the $n$-fold symmetrized direct 
product of $\Hs_1$ with itself. 
This space is spanned by vectors $\Phi_1 \otimes \cdots \otimes \Phi_n$,
where $\otimes$ denotes the symmetrized tensor product, representing
an $n$-particle state wherein the $k$-th particle is in the state
$\Phi_k \in \Hs_1$, $k = 1,\ldots n$. The representation $U_1(\Pid)$ induces 
a unitary representation $U_F(\Pid)$ on $\Hs_F$ by
\begin{equation} \label{covariant}
U_F(\lambda) \left( \Phi_1 \otimes \cdots \otimes \Phi_n \right) \coloneqq
U_1(\lambda) \Phi_1 \otimes \cdots \otimes U_1(\lambda) \Phi_n \, .
\end{equation}

     In interacting theories, the states in the corresponding physical
Hilbert space $\Hs$ do not have such an \textit{a priori} 
interpretation in physical terms, however. 
It is the primary goal of scattering 
theory to identify in $\Hs$ those vectors which describe, at asymptotic times, 
incoming, respectively outgoing, configurations of freely 
moving particles. Mathematically, this amounts to the construction of 
certain specific isometries (generalized M\o{}ller operators)
$\Omega^\In$ and $\Omega^\Out$ mapping $\Hs_F$
onto subspaces  $\Hs^{\In} \subset \Hs$ and  $\Hs^{\Out} \subset \Hs$, 
respectively, and intertwining the unitary actions
of the Poincar\'e group on $\Hs_F$ and $\Hs$. The resulting vectors 
\begin{equation} 
\left( \Phi_1 \otimes \cdots \otimes \Phi_n \right)^{\In / \Out}
\coloneqq \Omega^{\In / \Out} \,
\left( \Phi_1 \otimes \cdots \otimes \Phi_n \right) \in \Hs 
\end{equation}
are interpreted as incoming and outgoing particle configurations in 
scattering processes.

      If, in a theory, the equality $\Hs^{\In} =\Hs^{\Out}$ holds,
then every incoming scattering state evolves, after the
collision processes at finite times, into an outgoing scattering 
state. It is then physically meaningful to define on this space of 
states the scattering matrix, setting 
$S = \Omega^{\Out \, *} \, \Omega^{\In}$. Physical 
data such as collision cross sections can be derived from 
$S$  and the corresponding transition amplitudes 
$\langle \left( \Phi_1 \otimes \cdots \otimes \Phi_m \right)^{\In}, 
\left(\Phi^\prime_{1} \otimes \cdots \otimes \Phi^\prime_n\right)^{\Out} 
\rangle$,
  respectively, by a standard procedure. It should be noted, however, 
  that the physically mandatory equality of the spaces of
  incoming and outgoing scattering states and the more stringent
  condition that every state has an interpretation in these
  terms, \ie $\Hs = \Hs^{\In} =\Hs^{\Out}$ (asymptotic completeness),
  has not yet been established in most models of interest. This problem
  of asymptotic completeness will be discussed in Sect.\ \ref{completeness}.
  It has interesting consequences in the Haag-Kastler-Araki setting, such
  as the celebrated Bisognano-Wichmann property \cite{2Enc,DM19,Mu}. 
  Furthermore, any asymptotically complete quantum field theory
  can be presented on Fock space such that its underlying unitary
  representation of the Poincar\'e group is given by \eqref{covariant}.
  In other words, this representation does not contain
  information about the interaction by itself. It is 
  its action on the local  operators which matters. 

Before going into details, let us state 
the few physically motivated postulates entering into 
the analysis. As discussed, the point of departure is a family
of algebras $\As(\Os)$, more precisely a net, associated with 
the open subregions $\Os$ of Minkowski space and acting on $\Hs$.
Restricting attention to the case of Bosons, we may 
assume that this net is local in the sense that
if $\Os_1$ is spacelike separated from $\Os_2$, then all elements of
$\As(\Os_1)$ commute with all elements 
of $\As(\Os_2)$.\footnote{In the presence of Fermions, these 
algebras contain also fermionic operators which anti-commute.} 
This is the mathematical expression of the principle of 
Einstein causality.
The unitary representation $U$ of 
$\Pid$ acting on $\Hs$ is assumed to satisfy the 
relativistic  spectrum condition (positivity of energy in all
Lorentz frames) and, in the sense of equality of 
sets, $U(\lambda) \As(\Os) U(\lambda)^{-1} = \As(\lambda \Os)$
for all $\lambda \in \Pid$ and regions $\Os$,
where $\lambda \, \Os$ denotes the Poincar\'e transformed
region. It is also assumed that
the subspace of $U(\Pid)$-invariant vectors is spanned by a single
unit vector $\Omega$, representing the vacuum, which has 
the Reeh-Schlieder property, \ie 
each set of vectors $\As(\Os) \Omega$ is  
dense in $\Hs$. These standing assumptions
will subsequently be amended by further conditions concerning 
the particle content of the theory.

\section{Haag-Ruelle Theory} 

     Haag and Ruelle were the first to establish the 
     existence of scattering states within this general framework,
     cf.\ \cite{Jost};
     further substantial improvements are due to Araki and Hepp,
     cf.\ \cite{Ar}.
In all of these investigations, the arguments were given for 
quantum field theories with associated particles (in the Wigner sense)
which have strictly positive mass $m > 0$ and for which $m$ is an
isolated eigenvalue of the mass operator (upper and lower mass gap).
Moreover, it was assumed that states
{containing the corresponding particle with non-zero
probability} can be 
created from the vacuum by local operations. 
These assumptions allow only for theories with short range
interactions and particles carrying localizable charges.

     In view of these limitations, Haag-Ruelle theory has been developed in
a number of different directions. By now, the scattering theory
of massive particles is under complete control, including also
particles carrying {cone-localizable (gauge or topological) charges
in physical spacetime, and the limiting case of particles which are only 
localizable in wedge-shaped regions, bounded by two characteristic planes. 
Theories in low dimensions were also discussed, admitting particles
with exotic statistics (anyons, plektons). Due to 
page limitations}, these results must go without further 
mention; we refer the interested reader to the articles 
{\cite{BrMu, BuFr, Du18, 3Enc}}.
Theories of massless particles and of particles carrying 
charges of electric or magnetic type (infraparticles) will be 
discussed in subsequent sections.

     We outline here a generalization of 
Haag-Ruelle scattering theory presented in \cite{Dy}, which 
covers massive particles with localizable charges without 
relying on any further constraints on the shape of the
mass spectrum. 
In particular, the scattering of electrically neutral,
stable particles fulfilling a sharp dispersion law in the presence of
massless particles is included (\eg neutral atoms in their ground
states). Mathematically, this assumption can be 
expressed by the requirement that there exists a subspace 
$\Hs_1 \subset \Hs$ such that the restriction of $U(\Pid)$ to $\Hs_1$ 
is a representation of mass $m > 0$. 
We denote by $P_1$ the projection in $\Hs$ onto $\Hs_1$.

     To establish notation, let $\Os$ be a bounded spacetime region 
and let $A \in \As(\Os)$ be any operator such that 
$P_1 A \Omega \neq 0$. The existence of such localized (in brief, local) 
operators  amounts to the assumption that the particle carries a 
localizable charge. That the particle is stable, 
\ie completely decouples from the underlying continuum states, can be 
cast into a condition first stated by Herbst:
For all sufficiently small $\mu > 0$
\begin{equation} \label{herbst}
\Vert E_\mu (1 - P_1) A \Omega \Vert \leq c \, \mu^\eta \, ,
\end{equation}
for some constants $c, \eta > 0$, where $E_\mu$ is the projection 
onto the spectral subspace of the mass operator corresponding 
to spectrum in the interval $(m - \mu, m + \mu)$. In the case
originally considered by Haag and Ruelle, where $m$ is
isolated from the rest of the mass spectrum, this condition
is certainly satisfied. 

Setting $A(x) \coloneqq U(x) A U(x)^{-1}$,
where $U(x)$ is the unitary implementing the spacetime
translation\footnote{The velocity of light and 
Planck's constant are put equal to $1$ in what follows}
$x = (x_0, \vec{x})$, one  puts, for $t \neq 0$,
\begin{equation} \label{haagruelleoperators}
A_t(f) = \int \! d^{4} \! x \, g_t(x_0) f_{x_o} (\vec{x}) A(x) \, .
\end{equation}
Here $ x_0 \mapsto g_t(x_0) \coloneqq g \big( (x_0 - t)/|t|^\kappa \big) /
|t|^\kappa$ induces a time averaging about $t$,
$g$ being any test function which integrates
to $1$, whose Fourier transform has compact support, 
and $1/(1+ \eta) < \kappa < 1$ with $\eta$ as above. 
The Fourier transform of $f_{x_0}$ is given by 
$\widetilde{f_{x_0}}(\vec{p}) \coloneqq 
\widetilde{f}(\vec{p}) e^{-ix_0\omega(\vec{p})}$,
where $f$ is some test function on $\RR^3$ 
with $\widetilde{f}(\vec{p})$ having compact support, and  
$\omega(\vec{p}) = (\vec{p}^{\, 2} + m^2)^{1/2}$. 
Note that $(x_0, \vec{x}) \mapsto f_{x_0}(\vec{x})$ is
a solution of the Klein-Gordon equation of mass $m$.

    With these assumptions, it follows by a straightforward
application of the harmonic analysis of unitary groups that
in the sense of strong convergence
$A_t(f)\Omega \rightarrow P_1 A(f)\Omega$ 
and $A_{\, t}(f)^*\Omega \rightarrow 0$ as $t \rightarrow \pm\infty$, 
where $A(f) = \int \! d^3 \! x \, f (\vec{x}) A(0, \vec{x})$.
Hence, the {limits of the } operators $A_t(f)$ may be thought of as  
creation operators and their adjoints as 
annihilation operators. These operators are the basic ingredients 
in the construction of scattering states. Choosing local operators
$A_{\, k}$ as above and test functions $f^{(k)}$ with disjoint
compact supports in momentum space, $k = 1, \dots n$, the scattering
states are obtained as limits of the Haag-Ruelle approximants
\begin{equation} \label{approximations}
A_{1 \, t}(f^{(1)}) \cdots A_{n \, t}(f^{(n)}) \Omega \, .
\end{equation}
Roughly speaking, the operators 
$A_{k \, t}(f^{(k)})$ are localized in spacelike separated 
regions at asymptotic times $t$, due to the support properties 
of the Fourier transforms of the functions $f^{(k)}$. 
Hence they commute asymptotically because of locality and,
by the clustering properties of the vacuum state, the above vector becomes
a product of single-particle states. 
In order to prove convergence one proceeds, in analogy
to Cook's method in quantum mechanical scattering theory,
to the time derivatives,
\begin{equation} \label{derivative}
\begin{split}
& \partial_t \, A_{1 \, t}(f^{(1)}) \cdots A_{n \, t}(f^{(n)}) \Omega  \\ 
& =  \sum_{k \neq l}^{} \, A_{1 \, t}(f^{(1)}) \cdots 
[\partial_t  A_{k \, t}(f^{(k)}), A_{l \, t}(f^{(l)})] \cdots 
A_{n \, t} (f^{(n)}) \Omega  \\
& + \sum_{k} \, A_{1 \, t}(f^{(1)}) \cdots \overset{k}{\vee} 
\cdots A_{n \, t}(f^{(n)}) \, \partial_t   A_{k \, t} (f^{(k)}) \Omega \, ,  
\end{split}
\end{equation}
where $\overset{k}{\vee}$ denotes omission of $ A_{k \, t}(f^{(k)})$.
Employing techniques of Araki and Hepp, one can prove 
that the terms in the second line, involving 
commutators, decay rapidly in norm
as $t$ approaches infinity because of locality, as indicated above.
By applying condition (\ref{herbst}) and the fact that 
the vectors $\partial_t   A_{k \, t} (f^{(k)}) \Omega$ do
not have a component in the single-particle space 
$\Hs_1$, the terms in the third line 
can be shown to decay in norm like $|t|^{-\kappa(1 + \eta)}$.
Thus the norm of the vector (\ref{derivative}) is integrable
in $t$, implying the existence of the strong limits
\begin{equation}
\left( P_1 A_1(f^{(1)}) \Omega \otimes \cdots \otimes 
  P_1 A_n(f^{(n)}) \Omega  \right)^{\In / \Out} 
\coloneqq \lim_{t \rightarrow \mp \infty} \,
A_{1 \, t}(f^{(1)}) \cdots A_{n \, t}(f^{(n)}) \Omega \, .
\end{equation}
As indicated by the notation, these limits depend only on
the single-particle vectors $P_1 A_{k}(f^{(k)}) \Omega \in \Hs_1$,
$k = 1, \dots n$, but not on the {underlying specific} choice of
operators and test functions. In order to establish 
their Fock structure, one employs results 
on clustering properties of vacuum correlation functions
in theories without strictly positive minimal mass.
Using this, one can compute inner products of arbitrary asymptotic
states and verify that the maps
\begin{equation} \label{Fock}
\left( P_1 A_1(f^{(1)}) \Omega \otimes \cdots \otimes 
  P_1 A_n(f^{(n)}) \Omega  \right) \mapsto
\left( P_1 A_1(f^{(1)}) \Omega \otimes \cdots \otimes 
  P_1 A_n(f^{(n)}) \Omega \right)^{\In / \Out} 
\end{equation}
extend by linearity to isomorphisms $\Omega^{\In / \Out}$ from the 
Fock space $\Hs_F$ onto the subspaces $\Hs^{\In / \Out} \subset \Hs$
generated by the collision states. Moreover, 
the asymptotic states transform under 
the Poincar\'e transformations $U(\Pid)$ as
\begin{equation}
\begin{split}
& U(\lambda) \, \left( P_1 A_1(f^{(1)}) \Omega \otimes \cdots \otimes 
  P_1 A_n(f^{(n)}) \Omega  \right)^{\In / \Out}  \\ 
& = \left(U_1(\lambda) P_1 A_1(f^{(1)}) \Omega \otimes \cdots \otimes 
  U_1(\lambda) P_1 A_n(f^{(n)}) \Omega \right)^{\In / \Out} \, . 
\end{split}
\end{equation}
Thus the isomorphisms $\Omega^{\In / \Out}$ intertwine the action
of the Poincar\'e group on $\Hs_F$ and  $\Hs^{\In / \Out}$. 
We summarize these results, which are vital for the physical 
interpretation of the underlying theory, in the following theorem.

\begin{theorem} Consider a theory of a particle of mass $m > 0$ which
  satisfies the standing assumptions and the stability condition
  (\ref{herbst}). Then there exist canonical
  isometries  $\Omega^{\In / \Out}$, mapping the Fock space $\Hs_F$
  based on the single-particle space $\Hs_1$ onto subspaces
  \mbox{$\Hs^{\In / \Out} \subset \Hs$}
  of incoming and outgoing scattering states. Moreover,
  these isometries intertwine the action of the Poincar\'e transformations
  on the respective spaces.
\end{theorem}

     Since the scattering states have been identified with Fock space,
asymptotic creation and annihilation operators act on $\Hs^{\In/\Out}$ in
a natural manner. This point will be explained in the following 
section.

\section{LSZ Formalism}

    Prior to the results of Haag and Ruelle, an axiomatic approach 
to scattering theory was developed by Lehmann, Symanzik and Zimmermann
(LSZ), based on time-ordered vacuum expectation values of  
quantum fields \cite{LeSyZi}. The relative advantage of their approach 
with respect to Haag-Ruelle
theory is that useful reduction formulas for the $S$-matrix greatly 
facilitate computations, in particular in perturbation theory. 
Moreover these formulas are the starting point of general studies of the
momentum space analyticity properties of the $S$-matrix (dispersion
relations), {cf.\ \cite{4Enc}}. Within the present general 
setting, the LSZ method was established by Hepp \cite{He}.

    For simplicity of discussion, we consider again a single
particle type of mass $m > 0$ and integer spin $s$, subject to
condition (\ref{herbst}). According to the results of the preceding
section, one then can 
consistently define asymptotic creation operators
on the scattering states, setting
\begin{equation} 
\begin{split}
& A (f)^{\In / \Out} \, 
\left( P_1 A_1(f^{(1)}) \Omega \otimes \cdots \otimes 
  P_1 A_n(f^{(n)}) \Omega \right)^{\In / \Out}  \\
& \coloneqq \lim_{t \rightarrow \mp \infty} \,
A_t(f) \left( P_1 A_1(f^{(1)}) \Omega \otimes \cdots \otimes 
P_1 A_n(f^{(n)}) \Omega \right)^{\In / \Out}  \\
& = \left( P_1 A(f^{}) \Omega \otimes  P_1 A_1(f^{(1)}) 
\Omega \otimes \cdots \otimes P_1 A_n(f^{(n)}) \Omega \right)^{\In / \Out} \, .
\end{split}
\end{equation}
Similarly, one obtains the corresponding asymptotic annihilation operators, 
\begin{equation} 
\begin{split}
& A (f)^{\In / \Out \, *}\left( P_1 A_1(f^{(1)}) 
\Omega \otimes \cdots \otimes 
P_1 A_n(f^{(n)}) \Omega \right)^{\In / \Out}  \\
&  = \lim_{t \rightarrow \mp \infty} \,
 A_t (f)^* \left( P_1 A_1(f^{(1)}) \Omega \otimes \cdots \otimes 
  P_1 A_n(f^{(n)}) \Omega \right)^{\In / \Out} = 0 \, .  
\end{split}
\end{equation}
The preceding two equalities hold if the Fourier transforms of 
the functions $f, f^{(1)}, \dots f^{(n)}$ have
disjoint supports. We mention as an aside that,
by replacing the time averaging function $g$ in the definition
of $A_t(f)$ by a delta function, the above formulas still hold. 
But the convergence is then to be understood in the weak Hilbert
space topology. In this form the above relations were anticipated
by Lehmann, Symanzik and Zimmermann (asymptotic condition).

     It is straightforward to proceed from these relations
to reduction formulas. Let $B$ be any local operator. Then one has, in the 
sense of matrix elements between outgoing and incoming 
scattering states, 
\begin{equation} 
\begin{split}
  & B \, A (f)^{\In}  - A (f)^{\Out} 
\, B = \lim_{t \rightarrow \infty} 
\left( B \, A_{-t} (f) \, - A_t (f) \, B \right) \\
& = \lim_{t \rightarrow \infty} \left(
\int \! d^{4} \! x \, f_{-t}(x) 
B A(x) - \int \! d^{4} \! x \, f_{t} (x) A(x) B \right),
\end{split}
\end{equation}
where  $f_t(x) \coloneqq g_t(x_0) f_{x_0} (\vec{x})$.
Due to the (essential) support properties of $f_{\pm t}$, the
contributions to the latter two integrals arise, for asymptotic $t$,  
from spacetime points $x$ where the 
localization regions of $A(x)$ and $B$ have a negative timelike 
(first term), respectively positive timelike (second term) distance.
Thus one may proceed from the products of these operators
to the time-ordered products $T(B A(x))$, where $T(B A(x)) = A(x)B$
if the localization region of 
$A(x)$ lies in the future of that of $B$,
and $T(B A(x)) = B A(x)$ if it lies in {its} past. 
It is noteworthy that a precise definition of the time ordering for 
finite $x$ is irrelevant in the present context --- any reasonable 
interpolation between the above relations will do. 
Similarly, one can define time-ordered products for an 
arbitrary number of local operators. 
The preceding limit can then be recast into
\begin{equation}
= \lim_{t \rightarrow \infty} 
\int \! d^{4} \! x \, (f_{-t}(x) - f_{t}(x)) T(B A(x)) \, .
\end{equation}
The latter expression has a particularly simple form in
momentum space. Proceeding to the Fourier transforms of $f_{\pm t}$,  
{one obtains in the limit of large $t$ 
\begin{equation}
\big(\widetilde{f_{-t}}(p) - 
\widetilde{f_{t}}(p) \big) / \big(p_0 - \omega(\vec{p})\big) 
\longrightarrow 
- 2\pi i \widetilde{f}(\vec{p}) \ \delta(p_0 - \omega(\vec{p})) \, .
\end{equation}
Denoting by $T(B \widetilde{A}(p))$ the Fourier transform 
of $T(B A(x))$, this implies 
\begin{equation}
B \, A (f)^{\In}  - A (f)^{\Out} \, B =
- 2\pi i \int \! d^3 \! p \, \widetilde{f}(\vec{p})
\, \big(p_0 - \omega(\vec{p})\big) T(B \widetilde{A}(-p)) 
\Big|_{p_0 = \omega(\vec{p})} \, . 
\end{equation}
The restriction of  
$\big(p_0 - \omega(\vec{p})\big) T(B \widetilde{A}(-p))$
to the manifold $\{ p \in \RR^4 : p_0 = \omega(\vec{p}) \}$
(the ``mass shell'') is meaningful in the sense of 
distributions on $\RR^3$.} By the same token, one obtains 
\begin{equation}
A (f)^{\Out *}  B - B \, A (f)^{\In *} =
- 2\pi i \int \! d^3 \! p \, \overline{\widetilde{f}(\vec{p})} \
\big(p_0 - \omega(\vec{p})\big) T(\widetilde{A^*}(p) B) 
\Big|_{p_0 = \omega(\vec{p})} \, .  
\end{equation}

      Similar relations, involving an arbitrary number of 
asymptotic creation and annihilation operators, can be 
established by analogous considerations. Taking matrix elements of these 
relations in the vacuum state and recalling the action of the
asymptotic creation and annihilation operators on scattering 
states, one arrives at the following result, which is central in all
applications of scattering theory.

\begin{theorem}
Consider the theory of a particle of mass $m > 0$ subject to 
the conditions stated in the preceding sections and let  
$f^{(1)}, \dots, f^{(n)}$ be any family of test functions 
whose Fourier transforms have compact and non-overlapping support.
Then 
\begin{equation}
\begin{split}
& \big\langle \big( P_1 A_1(f^{(1)}) \Omega \otimes \cdots \otimes
P_1 A_k(f^{(k)}) \Omega \big)^{\Out},
 \big( P_1 A_{k+1} (f^{(k+1)}) \Omega \otimes \cdots \otimes
P_1 A_n(f^{(n)}) \Omega \big)^{\In} \big\rangle \\
& = (-2\pi i)^n \idotsint \! d^3 \! p_1 \cdots d^3 \! p_n \,
\overline{\widetilde{f^{(1)}}(\vec{p}_1)} \cdots 
\overline{\widetilde{f^{(k)}}(\vec{p}_k)}  \widetilde{f^{(k+1)}}(\vec{p}_{k+1}) 
\cdots \widetilde{f^{(n)}}(\vec{p}_n) \, \times \\
& \times \prod_{i=1}^n  \big( p_{i_0} - \omega(\vec{p}_i) \big) \,
\big\langle \Omega, 
T \big( \widetilde{A_1^*}(p_1) \cdots  \widetilde{A_k^*}(p_k) 
\widetilde{A_{k+1}}(-p_{k+1}) \cdots \widetilde{A_n}(-p_n) \Omega
\big) \big\rangle \Big|^{j=1,\dots n}_{p_{j_0} = \omega(\vec{p}_j)} \, ,
\end{split}
\end{equation}
in an obvious notation.
\end{theorem}

     Thus the kernels of the scattering amplitudes 
in momentum space are obtained by restricting the 
(by the factor $\prod_{i=1}^n \, ( p_{i_0} - \omega(\vec{p}_i) )$)
amputated Fourier transforms of the vacuum expectation values  
of the time-ordered products
to the positive and negative mass shells,
respectively. These are the famous LSZ reduction formulas, 
which provide a convenient link between the 
time-ordered (Green's) functions of 
a theory and its asymptotic particle interpretation.

\section{Asymptotic Particle Counters}  \label{counters}

     The preceding construction of scattering states applies to 
a significant class of theories; but even if one restricts attention to 
the case of massive particles, it does not cover all situations 
of physical interest. For an essential input in the construction
is the existence of local operators interpolating between the 
vacuum and the single-particle states. There may be no such operators 
at one's disposal, however, either because the particle in 
question carries a non-localizable charge, or because the given 
family of operators is too small. The latter case appears, for example,  
in {global and local 
gauge theories, where in general only the observables are fixed 
by the principle of gauge invariance}, and the physical particle 
content as well as the corresponding interpolating operators 
are not known from the outset. As observables create from
the vacuum only neutral states, the above construction of 
scattering states then fails if charged particles are present. 
Nevertheless, thinking in 
physical terms, one would expect that the observables contain all
relevant information in order to determine the features  
of scattering states, in particular their collision
cross section. That this is indeed the case was first shown by
Araki and Haag, cf.\ \cite{Ar}.

     In scattering experiments the measured data are provided by detectors
(\eg particle counters) and coincidence arrangements of detectors. Essential
features of detectors are their lack of response in the vacuum state
and their macroscopic localization. Hence, within the present mathematical
setting, a general detector is
represented by a positive operator $C$ on the physical Hilbert 
space $\Hs$ such that
$C \, \Omega = 0$. Because of the Reeh-Schlieder
Theorem, these conditions cannot be satisfied by local
operators. However, they can be fulfilled by ``almost local''
operators. Examples of such operators are
easy to produce, putting $C = L^*L$ with 
\begin{equation} \label{localizingoperator}
L = \int \! d^{4} \! x \, f(x) \, A(x) \, ,
\end{equation}
where $A$ is any local operator and $f$ any test function whose Fourier 
transform has compact support in the complement of the closed 
forward lightcone (and hence 
in the complement of the energy momentum spectrum of 
the theory). In view of the properties of $f$ and the
invariance of $\Omega$ under translations, it
follows that $C = L^*L$ annihilates the vacuum and 
can be approximated with arbitrary precision by local operators. 
The algebra generated by {finite sums and products
  of} these operators $C$ will be denoted by {$\Cs_0$}. 

   When preparing a scattering experiment, the first thing one 
must do 
with a detector is to calibrate it, \ie test its response to sources of 
single-particle states. Within the mathematical setting, this amounts to 
computing the matrix elements of $C$ in states $\Phi \in \Hs_1$:
\begin{equation}
\langle \Phi , C \, \Phi \rangle = \iint \! d^3 \! p \, d^3 \! q \,\,
\overline{{\Phi}(\vec{p}\,)} \, {\Phi}(\vec{q}\,) \,
\langle \vec{p} \, | C | \vec{q} \, \rangle \, . 
\end{equation}
{Note that this single-particle state may be charged.}
Here $\vec{p} \mapsto {\Phi}(\vec{p}\,)$ is the momentum space
wave function of $\Phi$,  
$\langle \, { \cdot } \, | C | \, { \cdot } \, \rangle$ is the
kernel of $C$ in the single-particle space $\Hs_1$, and we have omitted
(summations over) indices labeling internal degrees of
freedom of the particle, if any. The relevant information about
$C$ is encoded in its kernel. As a matter of fact, one 
only needs to know its restriction to the diagonal, 
$ \vec{p} \mapsto \langle \vec{p} \, | C | \vec{p} \, \rangle$. 
It is called the sensitivity function of $C$ 
and can be shown to be regular under quite general 
circumstances, cf.\ \cite{Ar,BuFr}.

   Given a state $\Psi \in \Hs$ for which the expectation 
value $\langle \Psi, C(x) \Psi \rangle$ differs 
significantly from $0$, one concludes that this state deviates 
from the vacuum in a region about $x$. For finite $x$, this does 
not mean, however, that $\Psi$ has a particle interpretation at $x$. 
For that spacetime 
point may be, {for example, the location of a collision center,
  where two or more particles come close together with
  non-vanishing probability.}
Yet if one proceeds to asymptotic times, one expects,
in view of the spreading of wave packets, that the
probability of finding two or more particles in the same 
spacetime region is dominated by the single-particle 
contributions. It is this physical insight which 
justifies the expectation that the detectors $C(x)$ become particle 
counters at asymptotic times. Accordingly, 
one considers for asymptotic $t$ the operators 
\begin{equation}
  C_{\, t}(h) \coloneqq \int \! d^3 \! x \, h(\vec{x}/t) \, C(t,\vec{x}) \, ,
  \label{AH}
\end{equation}
where $h$ is any test function on $\RR^3$. 
The role of the integral is to sum up all single-particle contributions  
with velocities in the support
of $h$ in order to compensate for the decreasing probability of finding 
such particles at asymptotic times $t$ about the localization center of the 
detector. That these ideas are consistent
was demonstrated by Araki and Haag, who
established the following result \cite{Ar}.

\begin{theorem}
Consider, as before, the theory of a massive particle. Let
$C^{(1)}, \dots, C^{(n)} \in {\Cs_0}$ be any family of detector operators and 
let $h^{(1)}, \dots, h^{(n)}$ be any family of test functions
on $\RR^3$. Then, for every state $\Psi^{\Out} \in \Hs^{\Out}$ of finite energy,
\begin{equation} \label{equ}
\begin{split}
& \lim_{t \rightarrow \infty} \,
\big\langle \Psi^{\Out},
C^{(1)}_{\, t}({h^{(1)}}) \cdots C^{(n)}_{\, t}({h^{(n)}}) \, 
\Psi^{\Out} \big\rangle \\
& \! \! = 
\idotsint \! d^3 \! p_1 \cdots  d^3 \! p_n \, 
\big\langle \! \Psi^{\Out}, \rho^{\Out}(\vec{p}_1) 
\cdots \rho^{\Out}(\vec{p}_n)  \Psi^{\Out} \big\rangle \, 
\prod_{k=1}^n h(\vec{p}_k/\omega(\vec{p}_k)) \, \langle \vec{p}_k | C^{(k)} |
\vec{p}_k \rangle \, . 
\end{split}
\end{equation}
{Here} $\rho^{\Out}(\vec{p}\,)$ is the momentum space density 
(the  product of creation and annihilation operators) of 
outgoing particles of momentum $\vec{p}$, and (summations over) possible  
indices labeling internal degrees of freedom of the particle
are omitted. An analogous relation holds for incoming scattering
states at negative asymptotic times.
\end{theorem}

     This result shows, first of all, that the scattering states
have indeed the desired interpretation with regard to the observables, 
as anticipated in the preceding sections. Since the
assertion holds for all scattering states of finite energy, one
may replace in the above theorem the outgoing 
scattering states by any state of finite energy, if 
the theory is asymptotically complete, \ie $\Hs = \Hs^\In = \Hs^\Out$.
Then choosing, in particular, any incoming scattering state and 
making use of the arbitrariness of the test functions
$h^{(k)}$ as well as the knowledge of the sensitivity functions of the
detector operators, one can compute the probability distributions 
of outgoing particle momenta in this state, and thereby 
the corresponding collision cross sections. 

    The question of how to construct certain specific 
incoming scattering states by using only local observables
was not settled by Araki and Haag, however. A general 
method to that effect was outlined in \cite{BuPoSt}. As a matter
of fact, for that method only the knowledge of states in the 
subspace of neutral states is required. Yet in this approach one would
need for the computation of, say, elastic collision cross sections of
charged particles the vacuum correlation functions involving 
at least eight local observables. This practical disadvantage of increased 
computational complexity of the method is offset by the conceptual 
advantage of making no appeal to quantities which are {\it a priori}
non-observable.  

  If one drops the assumption of asymptotic
  completeness and replaces the vector in equation \eqref{equ}
  by an arbitrary vector of bounded energy,
  one may still ask if the Araki-Haag detectors converge. 
  Such convergence results have been obtained by Dybalski and
  G\'erard in~\cite{DyG14, JK24}. The class of Araki-Haag detectors
  for which this convergence is under
  control is quite large: the union of their ranges coincides with
  the subspace of scattering states (with the vacuum omitted).
  One may hope that these results are a step towards proving
  asymptotic completeness, cf.\ Sect.~7. Let us mention as an aside
  that the limits of the Araki-Haag detectors correspond to the 
  asymptotic observables in many body quantum mechanics
  whose existence has been a vital step in proofs
  of asymptotic completeness~\mbox{\cite[Ch.\ 6.6]{DG97}}. 

\section{Massless particles and Huygens' principle}
   
    The preceding general methods of scattering theory apply only to
massive particles. Yet taking advantage of the salient fact that
massless particles always move with the speed of light, Buchholz
succeeded in establishing a scattering theory also for such particles.
Moreover, his arguments lead to a quantum version of
Huygens' principle \cite{Bu}.

     As in the case of massive particles, one assumes that there
is a subspace $\Hs_1 \subset \Hs$ corresponding to a representation 
of $U(\Pid)$ of mass $m=0$ and, for simplicity, integer helicity; 
moreover, there must exist local 
operators interpolating between the vacuum and the  
single-particle states. These assumptions cover, in particular, 
the important examples of the photon and of Goldstone particles. 
Picking any suitable local operator $A$ interpolating 
between $\Omega$ and some vector in $\Hs_1$, one sets,  
in analogy to (\ref{haagruelleoperators}),
\begin{equation}
A_t \coloneqq  
\int \! d^{4} \! x \, g_t(x_0) \,  \, 
(-1/2 \pi) \varepsilon (x_0) \, \delta (x_0^2 -
\vec{x}^{\, 2}) \, \partial_0 A (x) \, . 
\end{equation} 
Here $g_t(x_0) \coloneqq (1 / \ln |t|) \, g\big( (x_0 -t) / |\ln t| \big)$ 
with $g$ as in (\ref{haagruelleoperators}), and the solution of
the Klein-Gordon equation in (\ref{haagruelleoperators})
has been replaced by the fundamental solution of the wave equation
{(Pauli-Jordan function)};
furthermore, $\partial_0 A (x)$ denotes the derivative of $A(x)$ with
respect to $x_0$. Then, once again, 
the strong limit of $A_t \Omega$ as 
$t \rightarrow \pm\infty$ is $P_1 A \Omega$, with $P_1$ the 
projection onto~$\Hs_1$. 

In order to establish the convergence of $A_t$ 
as in the LSZ approach, one now uses the 
fact that these operators are, at asymptotic times $t$, localized 
in the complement of some forward, respectively backward, 
light cone. Because of locality, they therefore commute with
all operators which are localized in the interior
of the respective cones. More specifically, 
let $\Os \subset \RR^4$ be the localization region of $A$ and let 
$\Os_\pm \subset \RR^4$ be the two regions having a 
positive, respectively negative, timelike distance from all points in
$\Os$. Then, for any operator $B$ which is compactly 
localized in $\Os_\pm$, respectively, one obtains 
$
\lim_{t \rightarrow \pm \infty} \, A_t B \Omega =
\lim_{t \rightarrow \pm \infty} \, B A_t \Omega =
B P_1 A \Omega \, .
$
This relation establishes the existence of the limits 
\begin{equation}
A^{\In / \Out} = \lim_{t \rightarrow \mp  \infty} \, A_t
\end{equation}
on the (by the Reeh-Schlieder property) dense sets of 
vectors $\{ B \Omega : B \in \As(\Os_\mp) \} \subset \Hs$.
It requires some more detailed analysis to
prove that the limits have all of the properties of a (smeared) free
massless field, whose translates $x \mapsto A^{\In / \Out}(x)$
satisfy the wave equation and have 
c-number commutation relations. {}From these free fields one can
then proceed to asymptotic creation and annihilation operators  
and construct asymptotic Fock spaces 
$\Hs^{\In / \Out} \subset \Hs$ of massless particles
and a corresponding scattering matrix as in the massive case. 
The details of this construction can be found in \cite{Bu}, cf.\ 
also \cite{Haag}.

      It also follows from these arguments 
that the asymptotic fields $A^{\In / \Out}$ of massless particles
emanating from a region $\Os$, \ie for which 
the underlying interpolating operators $A$ are localized in $\Os$,
commute with all operators localized in $\Os_\mp$, respectively.
This result may be understood as an expression of Huygens' principle.
More precisely, denoting by  $\As^{\In / \Out} (\Os)$  
the algebras of bounded operators generated by {those} asymptotic
fields $A^{\In / \Out}$, respectively, one arrives at the following quantum
version of Huygens' principle.

\begin{theorem} Consider a theory of massless particles as described above
and let $\As^{\In / \Out} (\Os)$ be the algebras generated by 
massless asymptotic fields $A^{\In / \Out}$ with $A \in \As(\Os)$. Then
\begin{equation}
\As^{\In} (\Os) \subset \As(\Os_-)^\prime
\quad \mbox{and} \quad \As^{\Out} (\Os) \subset \As(\Os_+)^\prime \, .
\end{equation}
Here the prime denotes the set of bounded operators commuting 
with all elements of the respective algebras (\ie their 
commutants).
\end{theorem}

\section{Beyond Wigner's Concept of Particle} \label{infra}

      There is by now ample evidence that Wigner's concept of particle
is too narrow in order to cover all particle-like structures
appearing in quantum field theory. Examples are the partons which show
up in non-abelian gauge theories at very small spacetime scales as
constituents of hadrons, but which do not appear at large scales due
to the confining forces.  Their mathematical description requires a
quite different treatment \cite{Bu3}, which cannot be discussed here.
But even at large scales, Wigner's concept does not cover all
stable particle-like systems, the most prominent examples being
particles carrying an abelian gauge charge, such as the electron and
the proton, which are inevitably accompanied by infinite clouds of
(``on shell'') massless particles.

      The latter problem was discussed first by Schroer, who coined
the term \textit{infraparticle} for such systems \cite{Sch}.  Later, Buchholz
showed in full generality that, as a consequence of Gauss' law, pure
states with an abelian gauge charge can neither have a sharp mass nor
carry a unitary representation of the Lorentz group \cite{Bu4}, thereby
uncovering the simple origin of results found by explicit
computations, notably in quantum electrodynamics \cite{Ste}.  Thus one
is faced with the question of an appropriate mathematical
characterization of infraparticles which generalizes the concept of
particle invented by Wigner.  Some significant steps in this
direction were taken by Fr\"ohlich, Morchio and
Strocchi, who based a
definition of infraparticles on a detailed spectral analysis of the
energy-momentum operators.  For an account of these developments and
further references, cf.\ \cite{Haag}.
      
      We outline here an approach, originated by Buchholz,
which covers all stable particle-like structures appearing in quantum
field theory at asymptotic times. It is based on Dirac's idea of
improper particle states with sharp energy and momentum. In the
standard (rigged Hilbert space) approach to giving mathematical
meaning to these quantities one regards them as vector-valued
distributions, whereby one tacitly assumes that the improper states
can coherently be superimposed so as to yield normalizable
states. This assumption is valid in the case of Wigner
particles but fails in the case of infraparticles. A more adequate
method of converting the improper states into normalizable ones is
based on the idea of acting on them with suitable localizing
operators. In the case of quantum mechanics, one could take as a
localizing operator any sufficiently rapidly decreasing function of
the position operator. It would map the improper ``plane wave
states'' of sharp momentum into finitely localized states which
thereby become normalizable. In quantum mechanics, these two
approaches can be shown to be mathematically equivalent.  The
situation is different, however, in quantum field theory.

In quantum field theory, the appropriate localizing
operators $L$ are of the form  (\ref{localizingoperator}). 
They constitute a (non-closed) left ideal  $\Ls$
in the C*-algebra $\As$ generated by all local operators.
Improper particle states of sharp energy-momentum $p$ can then be
defined as linear maps $| \, \cdot \, \rangle_p : \Ls \rightarrow \Hs$
satisfying\footnote{It is instructive to (formally) replace here $L$ 
by the identity operator, making it clear that this relation indeed 
defines improper states of sharp energy-momentum.}
\begin{equation}
U(x) \, | L \rangle_p = e^{ipx} \, | L(x) \rangle_p \, , \quad L \in \Ls
\, .
\end{equation}
In theories of massive particles, one can always find localizing operators
$L \in \Ls$ such that their images $ | L \rangle_p \in \Hs$ 
are states with a sharp mass. This is the situation covered in  
Wigner's approach. In theories with long range forces there are, in general, 
no such operators, however, since the process of localization
inevitably leads to the production of low energy massless particles. 
Yet improper states of sharp momentum still exist in this
situation, thereby leading to a meaningful generalization of
Wigner's particle concept.

     That this characterization of particles covers all situations 
of physical interest can be justified in the general setting of 
relativistic quantum field theory as follows.
Picking $g_t$ as in (\ref{haagruelleoperators})
and any vector $\Psi \in \Hs$ with finite energy, one can show \cite{Bu2}
that the functionals $\rho_t$, $t \in \RR$, given by 
\begin{equation} \label{e.26}
\rho_t(L^* L) \coloneqq \int d^4 \! x \, g_t(x_0) \, 
\langle \Psi, (L^*L)(x) \Psi \rangle \, , \quad L \in \Ls \, ,
\end{equation}
are well defined and form an 
equicontinuous family with respect to a certain natural locally 
convex topology on the algebra 
$\Cs = \Ls^* \Ls$. This family of functionals therefore has, 
as $t \rightarrow \pm\infty$,  weak-*-limit points, denoted by $\sigma$. 
The functionals $\sigma$ are positive 
on $\Cs$ but not normalizable. (Technically speaking, they are
weights on the underlying algebra $\As$.) 
Any such $\sigma$ induces a positive semidefinite scalar product
on the left ideal $\Ls$ given by 
\begin{equation}
\langle L_1 \mid L_2 \rangle \coloneqq \sigma(L_1^* L_2) \, , \quad 
L_1,L_2 \in \Ls \, .
\end{equation}
After quotienting out elements of zero norm and taking the completion, 
one obtains a Hilbert space and a linear map 
$L \mapsto |L \rangle$
from $\Ls$ into that space. Moreover, the spacetime translations 
act on this space by a unitary representation 
satisfying the relativistic spectrum condition. 

      It is instructive to compute these 
functionals and maps in theories of massive particles.
Making use of relation (\ref{equ}) in Section~\ref{counters} one obtains,
with a slight change of notation,  
\begin{equation} 
\langle L_1 \mid L_2 \rangle = 
\int \! d\mu ({p}) \, \langle {p} \, | L_1^* L_2 | {p} \, \rangle \, ,
\end{equation}
where $\mu$ is a measure giving the probability density 
of finding at asymptotic times in state
$\Psi$ a particle of energy-momentum $p$. Once again,  
possible summations over different particle types and internal 
degrees of freedom have been omitted here. Thus, 
setting $|L \rangle_p \coloneqq L \, |p \rangle$, one concludes
that the map $L \mapsto | L \rangle$ can be decomposed
into a direct integral of 
improper particle states of sharp energy-momentum,
$| \, \cdot \, \rangle = \int^{\oplus} 
 d\mu ({p})^{1/2} \, | \, \cdot \, \rangle_p $.  
It is crucial that this result can also be established
without any \textit{a priori} input about the nature of 
the particle content of the theory \cite{BuPoSt, Po04}, thereby 
providing evidence of the universal nature of the concept
of improper particle states of sharp momentum, as outlined here. 

\begin{theorem}
Consider a relativistic quantum field theory satisfying the 
standing assumptions. Then the maps $L \mapsto |L \rangle$
defined above 
can be decomposed into improper particle states 
of sharp energy-momentum $p$,
\begin{equation} \label{decomposition}
| \, \cdot \, \rangle = \int^{\oplus} 
 d\mu ({p})^{1/2} \, | \, \cdot \, \rangle_p \, ,
\end{equation}
where $\mu$ is some measure depending on the 
{underlying state} $\Psi$ and the 
respective time limit taken.
\end{theorem}

In theories of Wigner particles {with equal
internal degrees of freedom, the GNS representations
induced by pure particle weights $|\,\cdot \,\rangle_p$
are mutually unitarily equivalent for different $p$
of the same mass}. This property
fails for infraparticles, whose ``plane wave states"
cannot be coherently superimposed, there are no corresponding
normalizable states. As was shown by Buchholz,
that is the case for electrically charged particle weights, being a
consequence of Gauss's law. 
The absence of normalizable single electron states was also established
in various semi-relativistic models, cf.\ \cite{5Enc}.

     Although a general scattering theory based on improper particle
states has not yet been developed, some progress has been made in
\cite{BuPoSt}. There it is outlined how inclusive collision cross sections 
of scattering states, where an undetermined number of low energy 
massless particles remains unobserved, can be defined in
the presence of long range forces, in spite of the fact that
a meaningful scattering matrix may not exist.

\section{Asymptotic completeness} \label{completeness}

      Whereas the description of the asymptotic particle 
features of any relativistic quantum field theory can be based  
on an arsenal of powerful methods, the question
of when such a theory has a complete particle interpretation 
remains open to date. The only 
non-trivial (interacting) exceptions are certain two dimensional
models with factorizing $S$-matrices \cite{3Enc} and
theories of wedge-localized particles in higher dimensions \cite{Du18, DuDy}; 
the latter examples are only of moderate interest, however, since they
do not exhibit local observables. This situation is in striking 
contrast to the case of quantum mechanics, where the problem
of asymptotic completeness has been completely settled \cite{DG97}. 

      One may trace the difficulties in quantum field
theory back to the possible formation of superselection 
sectors \cite{Haag} and the resulting complex particle structures.
{This feature does not appear 
in quantum mechanical systems, where the Hilbert space
of particle states, including possible bound states,
is fixed from the outset}.
Thus the first step in establishing a complete 
particle interpretation in a quantum field theory 
has to be the determination of its 
full particle content. Here the methods outlined
in the preceding section provide a systematic tool 
since they are based on particle counters,
  cf.\ \eqref{e.26}, by which 
the charged particle content can be extracted at large
times.

{}From the resulting 
data, obtained in this manner,
one must then reconstruct the full physical Hilbert space
of the theory comprising all superselection sectors. For theories
in which only massive particles appear, such a 
construction has been established in \cite{BuFr}.
The Hilbert spaces arising from the construction in theories 
of massive particles contain all 
scattering states. The question of completeness can then be
recast into the familiar problem of the unitarity of the scattering   
matrix. It is believed that  phase space (nuclearity) properties of the 
theory are of relevance here \cite{Haag}. This is, in particular,
due to the fact that phase space conditions exclude certain models
with pathological particle
structure, such as generalized  free fields with absolutely
continuous mass distribution. 
 
     However, in theories with long range forces, where a meaningful
scattering matrix may not exist, this strategy is bound to fail. 
Nonetheless, as in most high energy scattering experiments, only 
some very specific aspects of the particle interpretation are really
tested, one may think of other meaningful formulations of completeness.
The interpretation of most scattering experiments relies on 
the existence of conservation laws, such as those for energy
and momentum. If a state has a complete particle
interpretation, it ought to be possible to fully recover its 
energy, say, from its asymptotic particle content, 
\ie there should be no contributions to its total energy 
which do not manifest themselves asymptotically in the form of particles.
Now the mean energy-momentum of a state $\Psi \in \Hs$ is given by 
$\langle \Psi,  P  \Psi \rangle$, $P$ being the energy-momentum
operators, and the mean energy-momentum contained in its asymptotic
particle content is $\int \! d\mu (p) \, p$, where
$\mu$ is the measure appearing in the decomposition
(\ref{decomposition}). Hence, in case of a complete particle 
interpretation the following should hold:
\begin{equation} \label{completenesseq}
\langle \Psi,  P  \Psi \rangle = \int \! d\mu (p) \, p \, .
\end{equation}
Similar relations should also hold for other conserved 
quantities which can be attributed to particles, such as 
charge, spin \etc. It seems that such a weak condition
of asymptotic completeness suffices for a consistent 
interpretation of most scattering experiments. One may
conjecture that relation (\ref{completenesseq}) and its 
generalizations hold
in all theories admitting a local stress energy tensor
and local currents corresponding to the charges.


\begin{thebibliography}{99}
\bibitem[{\bf E}]{} The references marked with an E in front
  of their numbers are articles solicited for this encyclopedia.
  The titles may change.
\end{thebibliography}
\end{document}